\documentclass[prl,twocolumn,aps,amsmath,amssymb]{revtex4}

\usepackage{graphicx}% Include figure files
\usepackage{dcolumn}% Align table columns on decimal point
\usepackage{bm}% bold math
\usepackage{amsmath}
\usepackage{amsfonts}
\newcommand{\drawsquare}[2]{\hbox{%
\rule{#2pt}{#1pt}\hskip-#2pt%  left vertical
\rule{#1pt}{#2pt}\hskip-#1pt%  lower horizontal
\rule[#1pt]{#1pt}{#2pt}}\rule[#1pt]{#2pt}{#2pt}\hskip-#2pt%upper horizontal
\rule{#2pt}{#1pt}}% right vertical

\newcommand{\Yfund}{\raisebox{-.5pt}{\drawsquare{6.5}{0.4}}}%  fund
\newcommand{\Ysymm}{\Yfund\hskip-0.4pt%
                    \Yfund}%  symmetric second rank
\def\symm{\Ysymm}

\def\drawbox#1#2{\hrule height#2pt
        \hbox{\vrule width#2pt height#1pt \kern#1pt
              \vrule width#2pt}
              \hrule height#2pt}

\def\Asym#1#2{\vcenter{\vbox{\drawbox{#1}{#2}
              \kern-#2pt       % line up boxes
              \drawbox{#1}{#2}}}}

\def\asymm{\Asym{6.4}{0.3}}

\newcommand{\be}{\begin{eqnarray}}
\newcommand{\ee}{\end{eqnarray}}

\begin{document}

\title{Orientifold theory dynamics and symmetry breaking}

\author{Francesco {\sc Sannino}}
 \email{francesco.sannino@nbi.dk}
\affiliation{NORDITA,  Blegdamsvej
  17, DK-2100 Copenhagen \O, Denmark }
 \author{Kimmo {\sc Tuominen}}\email{kimmo.tuominen@phys.jyu.fi}
 \affiliation{Department of Physics,  P.O. Box 35,
FIN-40014 University of Jyv\"askyl\"a, Finland\\
Helsinki Institute of Physics, P.O. Box 64, FIN-00014 University
of Helsinki, Finland }

\begin{abstract}
We show that it is possible to construct
explicit models of electroweak symmetry breaking in which the number of techniflavors needed to
enter the conformal phase of the theory is small and weakly
dependent on the number of technicolors. Surprisingly, the minimal
model with {\it just} two (techni)flavors, together with a suitable
gauge dynamics, can be made almost conformal. The theories we consider are generalizations of orientifold type gauge theories, in which the fermions are in
either two index symmetric or antisymmetric representation of the gauge group, as the underlying
dynamics responsible for the spontaneous breaking of the
electroweak symmetry. We first study their phase diagram, and use the fact that specific sectors
of these theories can be mapped into supersymmetric Yang-Mills theory to strengthen our
results. This correspondence allows us also to have information on part of the nonperturbative
spectrum. We propose and investigate some explicit models while briefly exploring relevant phenomenological consequences. Our theories not only can be tested at the next collider experiments but, due to their simple structure, can also be studied via current lattice simulations.
\end{abstract}

\maketitle

Dynamical breaking of the electroweak symmetry due to underlying
strong gauge dynamics is a natural possibility \cite{TC}, which has been
intensively investigated. For a nice summary and references, see
\cite{Hill:2002ap}.
The goal of this paper is to show that it is possible to construct
explicit models of electroweak symmetry breaking in which the number of techniflavors needed to
enter the conformal phase of the theory is small and weakly
dependent on the number of technicolors. Surprisingly, the minimal
model with just two (techni)flavors, together with a suitable
gauge dynamics, can be made almost conformal.

The starting point and motivation for this study is the recent
argument \cite{Armoni:2004uu} that {\em non}-supersymmetric
Yang-Mills theories with a Dirac fermion either in the two
index symmetric or antisymmetric representation of the gauge group
are nonperturbatively equivalent to supersymmetric Yang-Mills
(SYM) theory at large $N$, so that exact results established in
SYM theory should hold also in these ``orientifold" theories.
The orientifold theories at finite $N$ were studied in
\cite{Sannino:2003xe}.
Interestingly, the two-index antisymmetric representation
for $N=3$ matches ordinary QCD. This observation was made long ago by
Corrigan and Ramond \cite{Corrigan:1979xf}.

In this paper we adapt the orientifold type field theories to
describe dynamical electroweak symmetry breaking.
In the following, we refer to fermions in the two-index
antisymmetric (symmetric) representation as {\it A-types}  ({\it
S-types}). By studying the conformal window of these theories we will show that S-type theories
provide a natural
framework for walking technicolor models with a small number of
techniflavors. The idea of using higher dimensional
representations of the gauge group for technicolor like theories
was also proposed and investigated by Eichten and Lane \cite{Lane:1989ej}. However, here we will
show that, thanks to the identification
of some sectors of these theories with SYM, we will be able to make new relevant predictions.

We then construct simple generalizations of the basic technicolor
models. We show that the minimal theory with just two S-type
technifermions is already near its conformal window. We also
construct the next to minimal theory with one family of Dirac
fermions, but with A-type gauge interactions rather than QCD-like
ones. We show that this theory is almost conformal when the
underlying gauge theory is $SU(4)$.

Thanks to the fact that certain sectors of these theories can be mapped into supersymmetric
Yang-Mills theory our prediction for the phase diagram of the two index symmetric theories is
much more robust than for ordinary non-supersymmetric theories. Besides, part of the
nonperturbative spectrum of these theories is known in an expansion in the inverse of the number
of colors. This allows us to suggest new phenomenological signatures for these theories. We also
comment on electroweak precision measurements.

%\subsection*{Phase Diagrams}

Consider $N_f$ Dirac flavors in the two-index symmetric
representation (S-types) of the $SU(N)$ gauge group.
To make the phenomenologically appealing properties of the theory
explicit, consider the beta-function
$\beta=-{\beta_0}\frac{g^3}{16\pi^2} -{\beta_1}\frac{g^5}{(16\pi^2
)^2} +... $, with
\begin{eqnarray}
\beta_0 &=& \frac{11}{3}N - \frac{2}{3}N_f\,(N\pm 2) \ ,  \\
\beta_1 &=&\frac{34}{3}N^2 - N_f \left(N\pm
2\right)\left[\frac{10}{3}N + \frac{2}{N}\left(N\mp
1\right)\left(N\pm 2\right)\right] \ , \nonumber
\end{eqnarray}
where the upper sign is for the two-index symmetric representation, while
the lower sign is for the two-index antisymmetric representation.
Note first that at infinite $N$ and with one Dirac flavor we recover
the super Yang-Mills beta function, and for a generic $N_f$ we
recover the beta function of the theory with $N_f$ adjoint Weyl
fermions. {}The theory is asymptotically free for
$N_f< \frac{11}{2} {N}/({N\pm 2})$.
At infinite $N$ {A-types} and {S-types} are indistinguishable and
$N_f$ must be $<5$ to have an asymptotically free theory
\cite{Armoni:2004uu}. This result is already very interesting for
phenomenology since it severely limits the number of possible
techniflavors. This is very different from the technicolor
theories with fermions in the fundamental representation of
the gauge group. In such a case the maximum allowed number of
flavors increases linearly with the number of colors: the maximum
of $N_f$ is equal to $11N/2$ for these theories.

At a finite number of colors, S-types and A-types are
distinguishable from each other.
{}For S-types asymptotic freedom is lost already for three flavors when $N=2$ or
when $N=3$, while the upper bound of $N_f=5$ is reached for $N=20$
and it does not change when $N$ is further increased.
In the theory with S-types, we therefore now know that the number
of flavors must be smaller than $5$ for the theory to yield chiral
symmetry breaking. This takes into account that there is also a
conformal window of size $N_f^c<N_f<5$, with the critical value
$N_f^c$ to be determined shortly. We will show that a theory with
two S-types is very close to the conformal window from $N=2$ and
up to a quite large $N$. {}For A-types the situation is quite opposite.

Some of the problems of the simplest technicolor models, such as
providing ordinary fermions with a mass, are
alleviated when considering new gauge dynamics in which the
coupling does not run with the scale but rather walks, i.e.
evolves very slowly \cite{Holdom:1981rm,{Yamawaki:1985zg},{Appelquist:an}, MY, GG}.
Achieving a slowly evolving coupling constant
usually requires a quite large number of fermions in the
fundamental representation of the gauge group, and the number of
fermions is expected to increase linearly with the number of
colors. The lowest number of fermions is obtained for two colors,
and the predicted number of flavors is around eight.
However, the {\bf S}-parameter remains still a problem even if one considers
nonperturbative corrections
\cite{Appelquist:1998xf,{Appelquist:1999dq}}.

\begin{figure}[t]
\includegraphics[width=6truecm,height=3.5truecm]{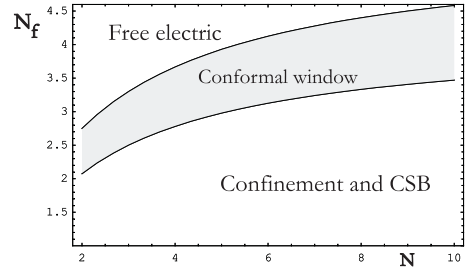}
\caption{Phase diagram as function of number of $N_f$ dirac
flavors and $N$ colors for fermions in the two-index symmetric
representation (S-types) of the gauge group. } \label{hawks}
\vskip -.5cm
\end{figure}

We will now show how to achieve walking in the theories considered here
with a low number of flavors. The critical value
of flavors, $N_f^c$, for the transition may be estimated by making
use of a perturbative expansion of the anomalous dimension of the
quark mass operator $\gamma$. At the first order in perturbation
theory, $\gamma$ is given by:
$\gamma = a_0\, \alpha$,
with $a_0=(3 C_2\left[R\right])/2\pi$ and $C_2\left[\symm\
,\asymm\right]=(N\pm 2)(N\mp1)/N$. We can now evaluate $\gamma$ at
the fixed point value of the coupling constant, which at two loops
in the beta function expansion is:
$\frac{\alpha^{\ast}}{4\pi}=- \frac{\beta_0}{\beta_1}$.
In Ref.~{\cite{ALM}}, it was noted that in the lowest (ladder)
order, the gap equation leads to the condition $\gamma (2-\gamma
)=1$ for chiral symmetry breaking. To all orders in perturbation
theory, this condition is gauge invariant (since $\gamma $ is
gauge invariant) and also equivalent to the condition $\gamma =1$
of Ref.~{\cite{CG}}. To any finite order in perturbation theory these
conditions are of course different. The condition $\gamma (2-\gamma )=1$
leads, in leading order, to the critical coupling
\begin{equation}
\alpha _{c}=\frac{\pi }{3C_{2}}\ ,  \label{critical}
\end{equation}
above which the ladder gap equation has a non-vanishing solution.
Using Eq.~(\ref{critical}) in $\alpha^{\ast}$, so that the
condition $\gamma <1$ is satisfied to leading order, leads to the conclusion that chiral
symmetry is restored for
\begin{equation}
N_{f}>N_{f}^{c}\simeq \frac{83N^3\pm 66N^2 - 132N}{20N^3\pm 55N^2
\mp 60}\ . \label{Ncrit}
\end{equation}
The phase diagram as a function of the number of colors and flavors
for the S-type case is presented in figure \ref{hawks}.
We find that for $N=2,3,4,5$ $N_f=2$ is already the highest
possible number of flavors before entering the conformal window.
Hence for these theories we expect a slowly evolving coupling
constant. {}We find that $N_f \geq 3$ for $N\geq 6$ but will
remain smaller than or equal to four for any $N$.

These estimates are based on the validity of the first
few terms in the perturbative expansion of the $\beta$-function.
We will provide in the next section another argument leading to the same prediction without
using any of the previous arguments.
The critical value of flavors increases with the number of colors
for the gauge theory with S-type matter: the limiting value is
$4.15$ at large $N$. Our results are consistent with the results
presented in \cite{Armoni:2004uu} only in the infinite $N$ limit.

\begin{figure}[htb]
%\vskip.5cm
\includegraphics[width=6truecm,height=3.5truecm]{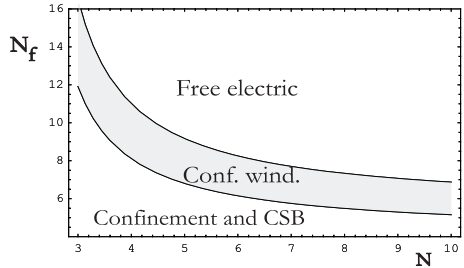}
\caption{Phase diagram as function of number of $N_f$ dirac
flavors and $N$ colors for fermions in the two-index antisymmetric
representation (A-types) of the gauge group. } \label{larks}
\vskip -.5cm
\end{figure}

The situation is different for the theory with A-type matter. As it is evident
from the associated phase diagram presented in figure \ref{larks},
the critical number of flavors increases when decreasing the
number of colors. The maximum value of $N_f=12$ is obtained for
$N=3$, i.e. standard QCD. {}We shall see that we can construct a
theory of eight A-type technifermions, corresponding to the one-family model,
which is already near the conformal window for $N=4$ and hence the
theory is another prototype walking technicolor model.

\subsection{Further constraints from super Yang-Mills}

A better understanding of the nonperturbative dynamics of these
theories can be obtained by exploiting their relation with
supersymmetric Yang--Mills theory.

The conformal window of the S-theory provided in the previous
section is more reliable than the correspondent one obtained for a
generic nonsupersymmetric theroy. This is so for two reasons: 1) The
S-theory looses asymptotic freedom quite soon. Already for $N_f=3$
or $N_f=4$ when $1<N\leq 5$ and for $N_f =5$ at large $N$. 2) The
$N_f=1$ sector of the theory (the orientifold field theory) at large
N is mapped in SYM which is known to confine. Hence we can exclude
the $N_f=1$ theory from being near conformal, up to $30\%$
corrections in the case of $N=3$. Clearly for $N > 3$ our prediction
is even more reliable. So we conclude that for $N=3$ the theory is
conformal or quasi-conformal already for $N_f=2$. We expect this to
hold also for the case $N=2$. The arguments used here about the
conformal window are {\it completely} independent on the
nonperturbative methods used above while not disagreeing with our
previous results.

The relation with super Yang--Mills can also be employed to deduce vital information about the
hadronic spectrum of nonsupersymmetric
theories and vice versa \cite{Sannino:2003xe,{Feo:2004mr}}. Indeed the $N_f=1$ bosonic sector of
the A/S-type theories, at large N, is mapped into the bosonic spectrum of SYM. In
\cite{Sannino:2003xe} and subsequently in \cite{Feo:2004mr} a number of relevant relations for
finite $N$ have been uncovered. The $N_f=1$ sector of a generic $N_f$ theory can then be
studied, up to $N_f$ corrections which can be taken into account. Since we are interested in
theories with $N_f=2$ the flavor corrections are small. We recall here some of the results
obtained in \cite{Sannino:2003xe} for the low lying bosonic hadronic states for the A and S type
theories with one flavor. These are the generalization of the QCD $\eta^{\prime}$ and associated
scalar $\sigma$, which in SYM are mapped in the pseudoscalar and scalar gluinoballs. In SYM
these states are degenerate and have a common mass. Introducing $1/N$ corrections one determines
\cite{Sannino:2003xe}:
\begin{eqnarray}
\frac{M_\sigma}{M_{\eta^{\prime}}} &\simeq& 1 \mp \frac{22}{9N}
\end{eqnarray}
where the upper (lower) sign refers to the S(A) theory. In the
previous formula another positive, but numerically small
\cite{Sannino:2003xe}, $1/N$ contribution proportional to the gluon
condensate has been neglected. The knowledge of the scalar sector of
these theories should be confronted with the elusive one in ordinary
QCD \cite{Black:1998wt,Sannino:1995ik,Harada:2003em}, and
consequently in walking as well as in ordinary technicolor theories.
If the present theories should ever emerge as the ones driving
electroweak symmetry breaking, we would predict that in the S(A)
type theory, even near the conformal window, the scalar companion of
the techni-$\eta^{\prime}$ would be lighter (heavier) than the
associated pseudoscalar meson. This suggests that the composite
Higgs, which is the chiral partner of the pion and typically lighter
than the $\eta^{\prime}$-partner, would also be lighter (heavier) in
the S(A) theory than in ordinary walking technicolor type theories.
Interestingly, if a light Higgs is discovered it would still be
consistent with a scenario of dynamical breaking of the electroweak
symmetry and perhaps induce a first order electroweak phase
transition in the early universe.

\subsection*{Model Building and Physical Predictions}
The minimal model with two fermions, which is almost conformal, has the following S-type
matter content and quantum numbers with respect to the Standard
Model:
\begin{eqnarray}
\left(%
\begin{array}{c}
  U^{\left\{ c_1,c_2\right\}} \\
  D^{\left\{ c_1,c_2 \right\}}\\
\end{array}%
\right)_{L} \quad {\rm with} \quad Y=0 \ ,
\end{eqnarray}
and $\left(U^{\left\{ c_1,c_2\right\}} , D^{\left\{
c_1,c_2\right\}} \right)_R$ with hypercharge $Y=(1,-1)$ and
$c_i=1,\ldots , N$ the gauge indices. {}This theory has symmetry
$SU(2)_L\times SU(2)_R\times U(1)$ for any $N\geq 3$ and it is
nearly conformal for $ 2\leq N \leq 5$.
Let us consider the cases $N=2$ and $N=3$ first. The two color theory has
a Witten anomaly \cite{Witten:fp}. To cure such an anomaly without introducing further unwanted
gauge anomalies one is
forced to introduce at least a new lepton family with half integer charge which should be
heavier than the ordinary top quark, but not much heavier. Another interesting point is that
the two index symmetric representation of $SU(2)$ is real, and hence the global classical
symmetry group is $SU(4)$ which breaks to $O(4)$. This leads to the appearance of nine
Goldstone bosons, of which three become the longitudinal components of the weak gauge bosons.
Hence, the low energy spectrum is expected
to contain six quasi goldstone bosons, which are expected to receive mass through
extended technicolor (ETC) interactions \cite{Hill:2002ap,ETC,Appelquist:2004mn}.
These predictions can be tested at LHC and will be further investigated in much more detail in a
near future.

The theory with three technicolors contains an even number of electroweak doublets, and hence
it is not subject to a Witten anomaly. Since the two index symmetric representation of $SU(3)$
is complex the flavor symmetry is $SU(2)_L\times SU(2)_R$. Only three goldstones emerge and
are absorbed in the longitudinal components of the weak vector bosons.
Based on the discussion reported in the previous section, we predict the Higgs to be lighter
than in ordinary walking technicolor theories.

Next we consider the one-family model in our framework. Here we
consider A-type technifermions:
\begin{eqnarray}
\left(%
\begin{array}{c}
  U^{\left[ c_1,c_2\right];C} \\
  D^{\left[ c_1,c_2 \right];C}\\
\end{array}%
\right)_{L} \quad {\rm with} \quad Y=y \ ,
\end{eqnarray}
with $C$ the ordinary color index.
\begin{eqnarray}
\left(%
\begin{array}{c}
  N^{\left[ c_1,c_2\right]} \\
  E^{\left[ c_1,c_2 \right]}\\
\end{array}%
\right)_{L} \quad {\rm with} \quad Y=-3\,y \ ,
\end{eqnarray}
and $\left(U^{\left[ c_1,c_2\right ];C} , D^{\left[
c_1,c_2\right];C} , N^{\left[ c_1,c_2\right ]},E^{\left[
c_1,c_2\right ]}\right)_R$ with hypercharge
$Y=(y+1,y-1,-3y+1,-3y-1)$ and generic $y$. The charge assignment
has been chosen such that the theory is free of gauge anomalies.
{}This theory has symmetry $SU(8)_L\times SU(8)_R\times U(1)$ for
any $N\geq 3$ and it is near conformal for $N =4$ with enhanced symmetry
$SU(16)$.

Electroweak precision measurements are an important test for any extension of the standard
model. {}
Clearly we need not to worry about the {\bf T} and {\bf U} parameters \cite{Peskin:1991sw} since
our theories naturally have built in custodial symmetry. The {\bf S} parameter whose experimental value is $-0.13\pm 0.11$ is generally a great problem for technicolor theories.
For S(A)-type models the perturbative contribution to the {\bf S} parameter is:
\begin{equation}
{\bf S}_{\rm pert.} (S/A) = {1 \over 6 \pi} \cdot \frac{N(N \pm 1)}{2}
\cdot \frac{N_{f}}{2} \ .
\end{equation}
Near-conformal dynamics leads to a further nonperturbative reduction in
the {\bf S} parameter \cite{Appelquist:1998xf}. $N_f$ is the number of techniflavors. It is
clear that only a small number of flavors
favors a small {\bf S} parameter, making the S-type theories with $N_f=2$ the best candidates of
walking-type dynamics not yet ruled out either at the perturbative or nonperturbative levels
\cite{Hong:2004td} by precision measurements. This situation should be confronted with the
ordinary walking technicolor dynamics which requires a large number of techniflavors to achieve
walking. Clearly this also shows that the A-type model we suggested is not favored by precision
measurements as investigated in much detail in the paper \cite{Hong:2004td} based on the present
work. We also expect for the A-type theories to have a Higgs heavier than the S-type one and
roughly of the same order than the ordinary techicolor type models.

Our findings suggest that, via the same dynamical mechanism, we may
be able to explain dynamical symmetry breaking of the electroweak
theory, and perhaps even the origin of $SU_L(2)$ as due to an almost
conformal theory. However, a more complete theory also able to
account for the mass generation of quarks and leptons is needed.

%%%%%%%%%%%%%%%%%%%%%%%%%%


\begin{thebibliography}{9}

\bibitem{TC}
S.~Weinberg,
%``Implications Of Dynamical Symmetry Breaking: An Addendum,''
Phys.\ Rev.\ D {\bf 19}, 1277 (1979);
%%CITATION = PHRVA,D19,1277;%%
L.~Susskind,
%``Dynamics Of Spontaneous Symmetry Breaking In The Weinberg-Salam Theory,''
Phys.\ Rev.\ D {\bf 20}, 2619 (1979).
%%CITATION = PHRVA,D20,2619;%%

%\cite{Hill:2002ap}
\bibitem{Hill:2002ap}
C.~T.~Hill and E.~H.~Simmons,
%``Strong dynamics and electroweak symmetry breaking,''
Phys.\ Rept.\  {\bf 381}, 235 (2003) [Erratum-ibid.\  {\bf 390},
553 (2004)].
%%CITATION = HEP-PH 0203079;%%

%\cite{Armoni:2004uu}
\bibitem{Armoni:2004uu}
A.~Armoni, M.~Shifman and G.~Veneziano,
%``From super-Yang-Mills theory to QCD: Planar equivalence and its
%implications,''
arXiv:hep-th/0403071;
%%CITATION = HEP-TH 0403071;%%
%\bibitem{Armoni:2003gp}
%A.~Armoni, M.~Shifman and G.~Veneziano,
%%``Exact results in non-supersymmetric large N orientifold field theories,''
Nucl.\ Phys.\ B {\bf 667}, 170 (2003), [hep-th/0302163];
%%%CITATION = HEP-TH 0302163;%%
%\bibitem{Armoni:2003fb}
%A.~Armoni, M.~Shifman and G.~Veneziano,
%``SUSY relics in one-flavor QCD from a new 1/N expansion,''
Phys.\ Rev.\ Lett.\  {\bf 91} (2003) 191601,
[arXiv:hep-th/0307097].
%%CITATION = HEP-TH 0307097;%%

%\cite{Sannino:2003xe}
\bibitem{Sannino:2003xe}
F.~Sannino and M.~Shifman,
%``Effective Lagrangians for orientifold theories,''
Phys.\ Rev.\ D {\bf 69}, 125004 (2004)
[arXiv:hep-th/0309252].
%%CITATION = HEP-TH 0309252;%%

%\cite{Corrigan:1979xf}
\bibitem{Corrigan:1979xf}
E.~Corrigan and P.~Ramond,
%``A Note On The Quark Content Of Large Color Groups,''
Phys.\ Lett.\ B {\bf 87}, 73 (1979).
%%CITATION = PHLTA,B87,73;%%

%\cite{Lane:1989ej}
\bibitem{Lane:1989ej}
K.~D.~Lane and E.~Eichten,
%``Two Scale Technicolor,''
Phys.\ Lett.\ B {\bf 222}, 274 (1989).
%%CITATION = PHLTA,B222,274;%% %\cite{Eichten:1979ah}
% \bibitem{Eichten:1979ah}
E.~Eichten and K.~D.~Lane,
%``Dynamical Breaking Of Weak Interaction Symmetries,''
Phys.\ Lett.\ B {\bf 90}, 125 (1980).
%%CITATION = PHLTA,B90,125;%%

%\cite{Holdom:1981rm}
\bibitem{Holdom:1981rm}
B.~Holdom,
%``Raising The Sideways Scale,''
Phys.\ Rev.\ D {\bf 24}, 1441 (1981).
%%CITATION = PHRVA,D24,1441;%%

%\cite{Yamawaki:1985zg}
\bibitem{Yamawaki:1985zg}
K.~Yamawaki, M.~Bando and K.~i.~Matumoto,
%``Scale Invariant Technicolor Model And A Technidilaton,''
Phys.\ Rev.\ Lett.\  {\bf 56}, 1335 (1986).
%%CITATION = PRLTA,56,1335;%%

%\cite{Appelquist:an}
\bibitem{Appelquist:an}
T.~W.~Appelquist, D.~Karabali and L.~C.~R.~Wijewardhana,
%``Chiral Hierarchies And The Flavor Changing Neutral Current Problem In
%Technicolor,''
Phys.\ Rev.\ Lett.\  {\bf 57}, 957 (1986);
%%CITATION = PRLTA,57,957;%%
%\cite{Appelquist:1998rb}
%\bibitem{ARTW}
T.~Appelquist, A.~Ratnaweera, J.~Terning and
L.~C.~R.~Wijewardhana,
%``The phase structure of an SU(N) gauge theory with N(f) flavors,''
Phys.\ Rev.\ D {\bf 58}, 105017 (1998).
%%CITATION = HEP-PH 9806472;%%

\bibitem{MY}  V.A.~Miransky and K.~Yamawaki, Phys. Rev. D{\bf 55}, 5051
(1997); erratum ibid. D{\bf 56}, 3768 (1997).

\bibitem{GG}
E.~Gardi and G.~Grunberg,
%``The conformal window in {QCD} and supersymmetric {QCD},''
JHEP {\bf 9903}, 024 (1999).
%%CITATION = HEP-TH 9810192;%%

%\cite{Appelquist:1998xf}
\bibitem{Appelquist:1998xf}
T.~Appelquist and F.~Sannino,
%``The physical spectrum of conformal SU(N) gauge theories,''
Phys.\ Rev.\ D {\bf 59}, 067702 (1999) [arXiv:hep-ph/9806409].
%%CITATION = HEP-PH 9806409;%%

%\cite{Appelquist:1999dq}
\bibitem{Appelquist:1999dq}
T.~Appelquist, P.~S.~Rodrigues da Silva and F.~Sannino,
%``Enhanced global symmetries and the chiral phase transition,''
Phys.\ Rev.\ D {\bf 60}, 116007 (1999) [arXiv:hep-ph/9906555].
%%CITATION = HEP-PH 9906555;%% \bibitem{Duan:2000dy}
Z.~y.~Duan, P.~S.~Rodrigues da Silva and F.~Sannino,
%``Enhanced global symmetry constraints on epsilon terms,''
Nucl.\ Phys.\ B {\bf 592}, 371 (2001) [arXiv:hep-ph/0001303].
%%CITATION = HEP-PH 0001303;%%

\bibitem{ALM}  T.~Appelquist, K.~Lane and U.~Mahanta, Phys. Rev. Lett.
{\bf 61} (1988), 1553.

\bibitem{CG}  A.~Cohen and H.~Georgi , Nucl. Phys. {\bf B314} (1989), 7.

%\cite{Feo:2004mr}
\bibitem{Feo:2004mr}
A.~Feo, P.~Merlatti and F.~Sannino,
%``Information on the super Yang-Mills spectrum,''
arXiv:hep-th/0408214. To Appear in Physical Review D.
%%CITATION = HEP-TH 0408214;%%

\bibitem{Black:1998wt}
D.~Black, A.~H.~Fariborz, F.~Sannino and J.~Schechter,
%``Putative light scalar nonet,''
Phys.\ Rev.\ D {\bf 59}, 074026 (1999) [arXiv:hep-ph/9808415].
%%CITATION = HEP-PH 9808415;%%

%\cite{Sannino:1995ik}
\bibitem{Sannino:1995ik}
F.~Sannino and J.~Schechter,
%``Exploring pi pi scattering in the 1/N(c) picture,''
Phys.\ Rev.\ D {\bf 52}, 96 (1995) [arXiv:hep-ph/9501417];
%%CITATION = HEP-PH 9501417;%%
%\cite{Harada:1995dc}
%\bibitem{Harada:1995dc}
M.~Harada, F.~Sannino and J.~Schechter,
%``Simple Description of Pion-Pion Scattering to 1 GeV,''
Phys.\ Rev.\ D {\bf 54}, 1991 (1996) [arXiv:hep-ph/9511335];
%%CITATION = HEP-PH 9511335;%%
%\cite{Harada:1996wr}
%\bibitem{Harada:1996wr}
M.~Harada, F.~Sannino and J.~Schechter,
%``Comment on *Confirmation of the sigma meson*,''
Phys.\ Rev.\ Lett.\  {\bf 78}, 1603 (1997) [arXiv:hep-ph/9609428];
%%CITATION = HEP-PH 9609428;%%
%\cite{Black:1998wt}
%\cite{Black:1998zc}
%\bibitem{Black:1998zc}
D.~Black, A.~H.~Fariborz, F.~Sannino and J.~Schechter,
%``Evidence for a scalar kappa(900) resonance in pi K scattering,''
Phys.\ Rev.\ D {\bf 58}, 054012 (1998) [arXiv:hep-ph/9804273].
%%CITATION = HEP-PH 9804273;%%

%\cite{Harada:2003em}
\bibitem{Harada:2003em}
M.~Harada, F.~Sannino and J.~Schechter,
%``Large N(c) and chiral dynamics,''
Phys.\ Rev.\ D {\bf 69}, 034005 (2004) [arXiv:hep-ph/0309206].
%%CITATION = HEP-PH 0309206;%%

%\cite{Witten:fp}
\bibitem{Witten:fp}
E.~Witten,
%``An SU(2) Anomaly,''
Phys.\ Lett.\ B {\bf 117}, 324 (1982).
%%CITATION = PHLTA,B117,324;%%

\bibitem{ETC}
S.~Dimopoulos and L.~Susskind,
%``Mass Without Scalars,''
Nucl.\ Phys.\ B {\bf 155}, 237 (1979);
%%CITATION = NUPHA,B155,237;%%
E.~Eichten and K.~D.~Lane,
%``Dynamical Breaking Of Weak Interaction Symmetries,''
Phys.\ Lett.\ B {\bf 90}, 125 (1980).
%%CITATION = PHLTA,B90,125;%%

%\cite{Appelquist:2004mn}
\bibitem{Appelquist:2004mn}
T.~Appelquist, M.~Piai and R.~Shrock,
%``Lepton dipole moments in extended technicolor models,''
arXiv:hep-ph/0401114 ; %``Fermion masses and mixing in extended technicolor models,''
 Phys.\ Rev.\ D {\bf 69}, 015002 (2004).
%%CITATION = HEP-PH 0308061;%%

%\cite{Peskin:1991sw}
\bibitem{Peskin:1991sw}
M.~E.~Peskin and T.~Takeuchi,
%``Estimation of oblique electroweak corrections,''
Phys.\ Rev.\ D {\bf 46}, 381 (1992).
%%CITATION = PHRVA,D46,381;%%

%\cite{Hong:2004td}
\bibitem{Hong:2004td}
D.~K.~Hong, S.~D.~H.~Hsu and F.~Sannino,
%``Composite Higgs from higher representations,''
Phys.\ Lett.\ B {\bf 597}, 89 (2004)
[arXiv:hep-ph/0406200].
%%CITATION = HEP-PH 0406200;%%



\end{thebibliography}
\end{document}